\documentclass[preprint,12pt]{elsarticle}
\usepackage{}

\usepackage{amsmath}
\usepackage{cases}
\usepackage{color}
\usepackage{hyperref}
\usepackage{amssymb}
\usepackage{graphicx}
\usepackage{graphics}
\usepackage{subfigure}
\usepackage{appendix}
\usepackage{textcomp}
\usepackage{bm}
\usepackage{url}

\makeatletter
  \newcommand\figcaption{\def\@captype{figure}\caption}
  \newcommand\tabcaption{\def\@captype{table}\caption}
\makeatother

\biboptions{numbers,sort&compress}

\begin{document}

\begin{center}

{\Large On the origin of intrinsic randomness of Rayleigh-B{\'e}nard turbulence}

\vspace{0.3cm}

 Zhiliang Lin$^1$, Lipo Wang$^2$ and Shijun Liao$^{1,3}$ \footnote{Corresponding author.  Email address: sjliao@sjtu.edu.cn}

\vspace{0.3cm}

$^1$ School of Naval Architecture, Ocean and Civil Engineering, Shanghai Jiaotong University, China;\\
$^2$ UM-SJTU Joint Institute, Shanghai Jiaotong University, Shanghai 200240, China;\\
$^3$ Ministry-of-Education Key Laboratory in Scientific and Engineering Computing, Shanghai, China

 \end{center}

\hspace{-0.6cm}{\bf Abstract}  {\em It is of broad interest to understand how the evolution of non-equilibrium systems can be triggered and the role played by external perturbations. A famous example is the origin of randomness in the laminar-turbulence transition, which is raised in the pipe flow experiment by Reynolds as a century old unresolved problem. Although there exist different hypotheses, it is widely believed that the randomness is ``intrinsic'', which, however, remains as an open question to be verified.  Simulating the modeled Rayleigh-B{\'e}nard convection system by means of the so-called clean numerical simulation (CNS) with negligible numerical noises that are smaller even than thermal fluctuation,  we verify that turbulence can be self-excited from the inherent thermal fluctuation, without any external disturbances, i.e. out of nothing.   This reveals a relationship between microscopic physical uncertainty and macroscopic randomness.   It is found that in physics the system nonlinearity functions as a channel for randomness information, and energy as well, to transport microscopic uncertainty toward large scales. Such scenario can generally be helpful to understand the various relevant phenomena. In methodology, compared with direct numerical simulation (DNS), CNS opens a new direction to investigate turbulent flows with largely improved accuracy and reliability.}

\vspace{0.3cm}

\hspace{-0.6cm}{\bf Key Words} turbulence, \sep chaos, \sep thermal fluctuation, \sep clean numerical simulation

\section{Introduction}
The evolution of non-equilibrium systems involves energy exchange through the system boundary with the surroundings. It is of broad interest to understand how such evolution can be triggered and what the function of external perturbation is. A famous example is the laminar-turbulent transition of the pipe flow first reported by Reynolds from his pioneering experiment~\cite{Reynolds-1883}. The continuous devoted efforts~\cite{Barkley-Nature-2015, Avila2011, Wu-PNAS-2015} have greatly enriched our understanding. For example, by measuring the puff decay and splitting time the critical Reynolds number in the 3D pipe flow can be numerically estimated (around 2040)~\cite{Avila2011,Orszag1970, Blackburn2004}. Besides, with the help of direct numerical simulation (DNS) with very fine resolution, both spatially and temporally, Wu et al. \cite{Wu-PNAS-2015} demonstrated the transition sensitivity to the pipe entrance condition. Physically laminar-turbulent transition is by nature closely relevant to disturbances, which can be both external and internal. It is widely believed that randomness is an intrinsic property in turbulence. However, till now our understanding of the origin and evolution mechanism of such intrinsic randomness is still unclear.

Numerically the Navier-Stokes (NS) equations can be solved by DNS with exactly the same initial/boundary conditions so as to exclude the external disturbances. Unfortunately, the sensitivity of nonlinear systems to numerical inaccuracy leads to severe deficiency of the solutions. As discovered by Lorenz, dynamic systems governed by the NS equations are essentially chaotic, i.e. due to the butterfly effect~\cite{Lorenz1963} the solutions have sensitive dependance, not only on the initial conditions (SDIC)~\cite{Lorenz1963} but also on numerical algorithms (SDNA)~\cite{Lorenz2006}. Because of the inevitable numerical noises, e.g. round-off  error and truncation error, the solution reliability of chaotic systems is very controversial~\cite{Yao2008}. Some spurious turbulence evolution cases from DNS have been reported in the literature~\cite{Wang2009, Pugachev2015}. In this sense DNS results are strongly numerical noise contaminated, although meaningful from the statistical point of view.
On the other hand, Wolfram  \cite{Wolfram2002}  mentioned  that  the Lorenz equations with the famous butterfly-effect are highly simplified and thus do not contain terms that represent viscous effects, and therefore he believes that these terms would tend to damp out small perturbations.

Fortunately, such kind of man-made uncertainty of numerical experiments can be well controlled by means of the clean numerical simulation (CNS)~\cite{Liao2009-Tellus,Wang2012,Liao2013-CSF,Liao2014,Li2014,Liao2015-IJBC}, which is based on an arbitrary-order Taylor series method (TSM) \cite{Barrio2005} and the arbitrary multiple-precision (MP) data \cite{MP}, together with solution verification check. For chaotic dynamic systems such as the well known three-body problem, the round-off error and truncation error can be largely reduced by CNS, even much less than the microscopic physical uncertainty due to wave-particle duality~\cite{Liao2013-CSF, Liao2014,Li2014, Liao2015-IJBC}, which is extremely small but inevitable. The obtained results ~\cite{Liao2013-CSF, Liao2014, Li2014, Liao2015-IJBC} indicate that macroscopic randomness in the three-body system can be self excited from the intrinsic microscopic physical uncertainty, at absence of any {\em external} disturbances. These convincing results are inspiring; however, more on the physics need to be explored to understand if such scenario can be generally valid in other more complicated systems (such as turbulence, although there are some tentative discussion on self randomization~\cite{Tsinober} and pattern formation~\cite{Cross}.

\begin{figure}[t]
    \begin{center}
        \begin{tabular}{cc}
            \includegraphics[width=10cm]{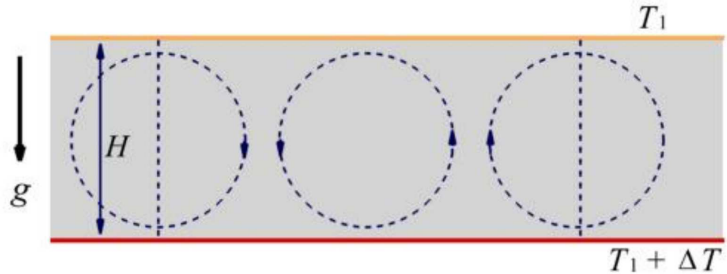}
        \end{tabular}
    \caption{{\bf Schematic representation of the Rayleigh-B{\'e}nard convective flow}. The two-dimensional incompressible fluid between two parallel free surfaces separated by $H$ obtains heat from the bottom boundary because of the prescribed constant temperature difference $\Delta T > 0$,  where $g$ is the gravity acceleration.} \label{Schematic}
    \end{center}
\end{figure}

\section{Methods}

\subsection{The governing equations and the spectral representation}

The numerical model considered here is a two-dimensional Rayleigh-B{\'e}nard (RB) system. As shown in Fig.~\ref{Schematic}, the incompressible fluid between two parallel free surfaces separated by $H$ obtains heat from the bottom boundary because of the prescribed constant temperature difference $\Delta T$, from which a reference velocity can be constructed as $\sqrt{g \alpha H\Delta T}$, where $g$ is the gravity acceleration and $\alpha$ is the thermal expansion coefficient of the fluid, respectively. This well-defined classic system has been extensively studied~\cite{Rayleigh1916,  Saltzman1962, Getling1998,  Malkus1954A,  Malkus1954B,  Roche2002, Niemela2006,  Ahlers2012, Zhou2013} either at its critical~\cite{Ahlers} or turbulent state~\cite{Grossman}. As described by Saltzman~\cite{Saltzman1962}, the corresponding non-dimensional governing equations in the form of  stream function $\psi$ with the Boussinesq approximation read
\begin{eqnarray}
\frac{\partial}{\partial t}\nabla^2\psi+\frac{\partial\left(\psi,
\nabla^2\psi\right)}{\partial(x, z)}-\frac{\partial
\theta}{\partial x} - {\cal C}_a\nabla^4\psi=0,\label{GEq01}\\
\frac{\partial\theta}{\partial t}+\frac{\partial\left(\psi,
\theta\right)}{\partial(x, z)} - \frac{\partial \psi}{\partial x}
- {\cal C}_b\nabla^2\theta =0,\label{GEq02}
\end{eqnarray}
where $\theta$ is the temperature departure from a linear variation background, $(x,z)$ are the horizontal and vertical spatial coordinates, $t$ denotes time,
$\nabla^2$ is the Laplace operator defined as $\nabla^4  = \nabla^2 \nabla^2$,
\[ \cfrac{\partial(a,b)}{\partial(x, z)}=  \cfrac{\partial a}{\partial x}\cfrac{\partial b}{\partial z}-\cfrac{\partial b}{\partial x}\cfrac{\partial a}{\partial z}  \]
 is the Jacobian operator, ${\cal C}_a = \sqrt{Pr/Ra}$ and ${\cal C}_b=1/\sqrt{Pr Ra}$ with the Rayleigh number $Ra=g\alpha H^3\Delta T/(\nu\kappa)$ and the Prandtl number $Pr=\nu/\kappa$, in which $\nu$ is the kinematic viscosity and $\kappa$ is the thermal diffusivity, respectively.
The free-slip boundary conditions at the upper and lower free surfaces read
\begin{eqnarray}
\frac{\partial\left(\psi, \nabla^2\psi\right)}{\partial(x,z)}=\frac{\partial\left(\psi, \theta\right)}{\partial(x, z)}=0.
\end{eqnarray}

Following Saltzman~\cite{Saltzman1962}, we express the stream function $\psi$ and temperature departure $\theta$ in the double Fourier expansion modes as
\begin{eqnarray}
\psi(x,z,t)=\sum_{m=-\infty}^{+\infty}\sum_{n=-\infty}^{+\infty}\Psi_{m,n}(t)
\exp\left[2\pi H
i\left(\frac{m}{L}x+\frac{n}{2H}z\right)\right],\label{PsiFExp}\\
\theta(x,z,t)=\sum_{m=-\infty}^{+\infty}\sum_{n=-\infty}^{+\infty}\Theta_{m,n}(t)
\exp\left[2\pi H
i\left(\frac{m}{L}x+\frac{n}{2H}z\right)\right],\label{ThetaFExp}
\end{eqnarray}
where $m,n$ are the wave numbers in the $x$ and $z$ directions, $\Psi_{m,n}(t)$ and $\Theta_{m,n}(t)$ denote the amplitudes of the stream function and temperature components with the wave numbers  $m$ and $n$, respectively. Substituting the above Fourier series into the original equations, we have the nonlinear dynamic system
\begin{eqnarray}
\dot\Psi_{m,n}(t)\!\!\!&=&\!\!\!\sum_{p=-\infty}^{+\infty}\sum_{q=-\infty}^{+\infty}
\frac{C_{m,n,p,q}\alpha_{p,q}^2}{\alpha_{m,n}^2}\Psi_{p,q}\Psi_{m-p,n-q}
-\frac{l^*m}{\alpha_{m,n}^2}\,i\,\Theta_{m,n}\nonumber\\
&-&{\cal C}_{a}\,\alpha_{m,n}^2\Psi_{m,n},\label{DotPsimn}\\
\dot\Theta_{m,n}(t)\!\!\!&=&\!\!\!-\sum_{p=-\infty}^{+\infty}\sum_{q=-\infty}^{+\infty}
C_{m,n,p,q}\Psi_{p,q}\Theta_{m-p,n-q} +l^*m\,i\,\Psi_{m,n}\nonumber\\
&-&{\cal C}_{b}\,\alpha_{m,n}^2\Theta_{m,n},\label{DotThetamn}
\end{eqnarray}
where $C_{m,n,p,q}=l^*h^*(mq-np)$,  $l^*=2\pi H/L$, $h^*=\pi$ and $\alpha_{m,m}^2=(l^{*2}m^2+h^{*2}n^2)$.
Write
\begin{eqnarray}
\Psi_{m,n}=\Psi_{1,m,n} -
i\,\Psi_{2,m,n},\;\;\;\;\Theta_{m,n}=\Theta_{1,m,n} -
i\,\Theta_{2,m,n}, \label{CmplxType}
\end{eqnarray}
with the definitions
\[ \Psi_{1,m,n} = \Psi_{1,-m,-n}, \Psi_{2,m,n} = -\Psi_{2,-m,-n},\]
\[ \Theta_{1,m,n} = \Theta_{1,-m,-n}, \Theta_{2,m,n}=-\Theta_{2,-m,-n}.\]
It thus yields the following set of coupled nonlinear differential equations
\begin{small}
\begin{eqnarray}
\dot\Psi_{1,m,n}\!\!\!&=&\!\!\!\sum_{p=-\infty}^{+\infty}\sum_{q=-\infty}^{+\infty}
\frac{C_{m,n,p,q}\alpha_{p,q}^2}{\alpha_{m,n}^2}\Big(\Psi_{1,p,q}\Psi_{1,m-p,n-q}
-\Psi_{2,p,q}\Psi_{2,m-p,n-q}\Big)\nonumber\\
\!\!\!&-&\!\!\!\frac{l^*m}{\alpha_{m,n}^2}\Theta_{2,m,n}-{\cal
C}_{a}\,\alpha_{m,n}^2\Psi_{1,m,n},\label{DotPsi01mn}\\
\dot\Psi_{2,m,n}\!\!\!&=&\!\!\!\sum_{p=-\infty}^{+\infty}\sum_{q=-\infty}^{+\infty}
\frac{C_{m,n,p,q}\alpha_{p,q}^2}{\alpha_{m,n}^2}\Big(\Psi_{1,p,q}\Psi_{2,m-p,n-q}
+\Psi_{2,p,q}\Psi_{1,m-p,n-q}\Big)\nonumber\\
\!\!\!&+&\!\!\!\frac{l^*m}{\alpha_{m,n}^2}\Theta_{1,m,n}-{\cal
C}_{a}\,\alpha_{m,n}^2\Psi_{2,m,n},\label{DotPsi02mn}\\
\dot\Theta_{1,m,n}\!\!\!&=&\!\!\!-\sum_{p=-\infty}^{+\infty}\sum_{q=-\infty}^{+\infty}
C_{m,n,p,q}\Big(\Psi_{1,p,q}\Theta_{1,m-p,n-q}
-\Psi_{2,p,q}\Theta_{2,m-p,n-q}\Big)\nonumber\\
\!\!\!&+&\!\!\!l^*m\Psi_{2,m,n}-{\cal
C}_{b}\,\alpha_{m,n}^2\Theta_{1,m,n},\label{DotTheta01mn}\\
\dot\Theta_{2,m,n}\!\!\!&=&\!\!\!-\sum_{p=-\infty}^{+\infty}\sum_{q=-\infty}^{+\infty}
C_{m,n,p,q}\Big(\Psi_{1,p,q}\Theta_{2,m-p,n-q}
+\Psi_{2,p,q}\Theta_{1,m-p,n-q}\Big)\nonumber\\
\!\!\!&-&\!\!\!l^*m\Psi_{1,m,n}-{\cal
C}_{b}\,\alpha_{m,n}^2\Theta_{2,m,n}. \label{DotTheta02mn}
\end{eqnarray}
\end{small}

The free-slip boundary condition implies
\begin{eqnarray}
\Psi_{1,m,n} = -\Psi_{1,m,-n}=-\Psi_{1,-m,n}, \\
\Psi_{2,m,n}  =  -\Psi_{2,m,-n} = \Psi_{2,-m,n},\\
\Theta_{1,m,n} =-\Theta_{1,m,-n}=-\Theta_{1,-m,n}, \\
\Theta_{2,m,n}=-\Theta_{2,m,-n}=\Theta_{2,-m,n}, \label{ThetaPro04}
\end{eqnarray}
with
\begin{eqnarray}
\Psi_{1,0,n}=\Theta_{1,0,n}=\Psi_{1,m,0}=\Psi_{2,m,0}
=\Theta_{1,m,0}=\Theta_{2,m,0}=0
\end{eqnarray}
at $z=0$ and $z=1$. For more details, please refer  to  Saltzman \cite{Saltzman1962}.

Numerically, only a finite number of wave numbers can be considered, i.e. $|m|\leq M, |p|\leq M$ and $|n|\leq N$, $|q|\leq N$. In principal, the turbulence physics can be well described if the mode numbers $M$ and $N$ are large enough, the same rules for DNS as well~\cite{Orszag1970, Blackburn2004, Ashley2009}. For the present Rayleigh-B{\'e}nard flow with $Ra = 10^7$, $M=N=127$ is large enough to investigate the laminar-turbulent transition.

It should be emphasized that the above nonlinear dynamic system might evolve to be chaotic and the numerical behaviors might be influenced by the butterfly-effect, i.e. the sensitive dependence on the initial conditions (SDIC) in which a small change of this deterministic nonlinear system results in large difference in a later state~\cite{Lorenz1963, Lorenz2006}. Therefore this dynamic system might be very sensitive to numerical noises, which could evolve exponentially with time~\cite{Lorenz1963, Lorenz2006}. To avoid the loss of accuracy, the conventional fast  Fourier transform method is {\it not} used here for the nonlinear terms.  This is very different from DNS~\cite{Orszag1970, Blackburn2004, Ashley2009}.  However, the computational cost need to increase largely.\\

\subsection{Thermal fluctuation as the initial random condition}

The thermal fluctuation plays an important role on hydrodynamic instability   \cite{Wu1995, Ahlers2003}.   Recently, Wang et al. \cite{Wang-PNAS-2015} investigated the instability of the two-dimensional Poiseuille flow via DNS by considering the evolution from the laminar state under the action of different initial Gaussian white noise at the macroscopic level. In the present work, we use the Gaussian white noise as the initial condition of the laminar flow as well. However, unlike  Wang et al. \cite{Wang-PNAS-2015}, the Gaussian white noises is set here as the thermal fluctuation at the micro-level, which is physically inevitable with clear meaning. For the studied cases, the fluid is water at the room temperature of 20$^{o}$C, the  standard deviations for the temperature and velocity field can be estimated from statistical mechanics~\cite{Khinchin,  Gorodetsky2004, Landau} as $\sigma_T=10^{-10}$ and $\sigma_u=10^{-9}$, respectively.

To ensure the solution accuracy, it requires that the numerical noises must be even less than the thermal fluctuation in a long enough time interval for the onset of turbulence, which is rather difficult to achieve for the chaotic system under consideration. Fortunately, the clean numerical simulation (CNS) makes it possible to attack this numerical challenge~\cite{Liao2009-Tellus, Liao2013-CSF, Liao2014, Liao2015-IJBC} in the way described below.

\subsection{The clean numerical simulation (CNS)}

Due to the famous butterfly-effect, chaotic dynamic systems have sensitive dependance not only on the initial conditions (SDIC)~\cite{Lorenz1963}  but  also  on  numerical  algorithms (SDNA)~\cite{Lorenz2006}. Unfortunately, numerical noises such as round-off error and truncation error are inevitable in practice, which make the convergent numerical simulations of chaotic systems rather difficult to obtain in a desired (finite but long enough) time interval. This challenge leads to intense arguments on the reliability and feasibility of numerical simulations of chaos.  It is even  believed  that ``all chaotic responses are simply numerical noise and have nothing to do with the solutions of differential equations''~\cite{Yao2008, Lorenz2008}. The Lorenz equations \cite{Lorenz1963} are the much simplified model of the Navier-Stokes equations, which suggests that dynamic systems related to the Navier-Stokes equations should be sensitive to numerical noises as well. Indeed, some {\em spurious} and non-physical evolutions of turbulence from DNS have currently been reported~\cite{Wang2009, Pugachev2015}, which originate either from round-off error or dependence upon the time step size.   Currently,  Hoovers \cite{Hoover2015} applied two symplectic and five Runge-Kutta integrators to investigate a chaotic Hamiltonian system and found that all of these schemes can {\em not} gain convergent trajectories.  Therefore it is necessary to develop a numerical technique to obtain convergent and reliable simulation results of chaotic dynamic systems in a finite but long enough time interval.

The so-called clean numerical simulation (CNS)~\cite{Liao2009-Tellus, Liao2013-CSF, Liao2014,Li2014,Liao2015-IJBC} was developed recently for this purpose. CNS is based on the Taylor series method~\cite{Corliss1982,Barrio2005} at {\em arbitrary} order (in time) and data in {\em arbitrary} precision~\cite{MP}, together with a solution verification in the temporal domain. The Taylor series method has an advantage that its formula at an arbitrarily high order can be easily expressed and analyzed to deduce truncation error to a required level. Moreover, the multiple-precision (MP) data \cite{MP} is used here to control the round-off error to a required level in CNS. A remarkable example of the MP data application is to calculate the value of $\pi$ to millions of digit numbers.

In 2009, Liao~\cite{Liao2009-Tellus} first successfully implemented CNS to obtain a convergent chaotic solution of the Lorenz equation in the time interval $[0,1000]$, with 400th-order Taylor series and 800-digit MP data. The reliability of this CNS result has been confirmed~\cite{Wang2012} by CNS with 1000th-order Taylor series and 2100-digit MP data in a longer time interval [0,2500]. Currently, using 1200 CPUs at the National Supercomputer TH-A1 (in Tianjin, China) and a parallel CNS algorithm with a 3500th-order Taylor expansion and 4180-digit MP data, Liao and Wang~\cite{Liao2014} have successfully obtained, for the first time, a convergent and reliable solution of the Lorenz equation in a rather long interval [0,10000], which is several hundred times longer than those from the traditional numerical algorithms (such as the Runge-Kutta method). This brand-new simulation result, never reported in open literature before, provides us a numerical benchmark for mathematically reliable long-term prediction of chaos.   The instability of some currently reported periodic solutions of three-body system was also investigated by means of the CNS \cite{Li2014}.    In addition, the evolution of a chaotic three-body system with inherent uncertainty of the initial positions at the micro-level have been reliably simulated by CNS in a long enough time interval~\cite{Liao2015-IJBC}.   Besides,  it is found that, unlike the symplectic integrators,  the CNS can give accurate trajectories of chaotic Hamiltonian systems in a long interval \cite{Li2016-A}.   Furthermore, it is currently reported that the numerical noises even have a significant influence on statistics of chaotic dynamic systems in non-equilibrium \cite{Li2016-B}.

Similarly the convergent and reliable solution of the dynamic system (9)-(12) can be gained numerically by CNS as well. Let $\Delta t$ denote the time increment and  $f^{(j)}$ the value of $f(t)$ at $t=j\Delta t$. The $P$th-order Taylor series of $\Psi_{i,m,n}$ and $\Theta_{i,m,n}$ are expressed as
\begin{eqnarray}
\Psi_{i,m,n}^{(j+1)}=\Psi_{i,m,n}(t_j+\Delta
t)=\Psi_{i,m,n}^{(j)}+\sum_{k=1}^P\beta_{i,m,n}^{j,k}\,(\Delta
t)^k,\label{TaySerOfPsi}\\
\Theta_{i,m,n}^{(j+1)}=\Theta_{i,m,n}(t_j+\Delta
t)=\Theta_{i,m,n}^{(j)}+\sum_{k=1}^P\gamma_{i,m,n}^{j,k}\,(\Delta
t)^k,\label{TaySerOfTht}
\end{eqnarray}
where
\begin{small}
\begin{eqnarray}
&& \beta_{1,m,n}^{j,k+1} \nonumber \\
&=& \left(\sum_{p=-M}^{M}\sum_{q=-N}^{N}C_{m,n,p,q}
\frac{\alpha_{p,q}^2}{\alpha_{m,n}^2}\sum_{l=0}^{k}\left[\beta_{1,p,q}^{j,l}\beta_{1,m-p,n-q}^{j,k-l}
-\beta_{2,p,q}^{j,l}\beta_{2,m-p,n-q}^{j,k-l}\right]\right.  \nonumber\\
&& \left.-\frac{l^*
m}{\alpha_{m,n}^2}\gamma_{2,m,n}^{j,k}  -{\cal
C}_{a}\,\alpha_{m,n}^2\beta_{1,m,n}^{j,k}\right)/(1+k),\\
&& \beta_{2,m,n}^{j,k+1}  \nonumber \\
& = &\left(\sum_{p=-M}^{M}\sum_{q=-N}^{M}C_{m,n,p,q}
\frac{\alpha_{p,q}^2}{\alpha_{m,n}^2}\sum_{l=0}^{k}\left[\beta_{1,p,q}^{j,l}\beta_{2,m-p,n-q}^{j,k-l}
+\beta_{2,p,q}^{j,l}  \beta_{1,m-p,n-q}^{j,k-l} \right]  \right.    \nonumber\\
&& \left.+\frac{l^* m}{\alpha_{m,n}^2}\gamma_{1,m,n}^{j,k} - {\cal C}_{a}\, \alpha_{m,n}^2\beta_{2,m,n}^{j,k} \right)/(1+k), \\
&& \gamma_{1,m,n}^{j,k+1} \nonumber\\
&=&\left(-\sum_{p=-M}^{M}\sum_{q=-N}^{+N}C_{m,n,p,q}
\sum_{l=0}^{k}\left[\beta_{1,p,q}^{j,l}\gamma_{1,m-p,n-q}^{j,k-l}-\beta_{2,p,q}^{j,l}\gamma_{2,m-p,n-q}^{j,k-l}\right]\right.\;\;\;\;\;\;\nonumber\\
&&  \left.+ l^* m\beta_{2,m,n}^{j,k} -{\cal C}_{b}\,
 \alpha_{m,n} \gamma_{1,m,n}^{j,k}\right)/(1+k),\\
&& \gamma_{2,m,n}^{j,k+1}  \nonumber  \\
&=& \left(-\sum_{p=-M}^{M}\sum_{q=-N}^{N}C_{m,n,p,q}
\sum_{l=0}^{k}\left[\beta_{1,p,q}^{j,l}\gamma_{2,m-p,n-q}^{j,k-l}+\beta_{2,p,q}^{j,l}\gamma_{1,m-p,n-q}^{j,k-l}\right]\right.\;\;\;\;\;\;\nonumber\\
&&  \left. - l^* m\beta_{1,m,n}^{j,k}  {\cal C}_{b}\, \alpha_{m,n}^2\gamma_{2,m,n}^{j,k}\right)/(1+k).
\end{eqnarray}
\end{small}

Here
\begin{eqnarray}
\beta_{1,m,n}^{j,0}=\Psi_{1,m,n}^{(j)}=\Psi_{1,m,n}(t_j),  \\
\beta_{2,m,n}^{j,0}=\Psi_{2,m,n}^{(j)}=\Psi_{2,m,n}(t_j),\\
\gamma_{1,m,n}^{j,0}=\Theta_{1,m,n}^{(j)}=\Theta_{1,m,n}(t_j),\\
\gamma_{2,m,n}^{j,0}=\Theta_{2,m,n}^{(j)}=\Theta_{2,m,n}(t_j).
\end{eqnarray}

\section{Results}

\subsection{Modeling and numerical simulation}
To attack the randomness physics in turbulence, we focus on a two-dimensional Rayleigh-B\'{e}nard (RB) model system\cite{Rayleigh1916, Saltzman1962, Getling1998}, as shown in Fig.~1.  Without loss of generality, we consider the case with the aspect ratio $\Gamma = L/H = 2\sqrt{2}$,  Prandtl number $Pr = 6.8$  (water) and Rayleigh number $Ra =10^{7}$, corresponding to a linearly unstable case.

Similarly as in the 2D Poiseuille flow instability analysis by Wang et al.~\cite{Wang-PNAS-2015}, we formulate the inevitable thermal fluctuation as Gaussian white noise. It need to mention here that such Gaussian white noise is more than a random input, but physically meaningful to represent the thermal fluctuation in the nonlinear fluid system. To ensure numerical accuracy, CNS is adopted to simulate the evolution of the micro-level thermal fluctuation that is much less than the numerical noises of DNS. We set here the double Fourier expansion modes as $M = N = 127$, the multiple-precision data in 100-digits, the 10th-order ($P=10$) of the truncated Taylor series in time with the step size $\Delta t = 5 \times 10^{-3}$.

\begin{figure}[t]
    \begin{center}
        \begin{tabular}{cc}
            \includegraphics[width=0.75\textwidth]{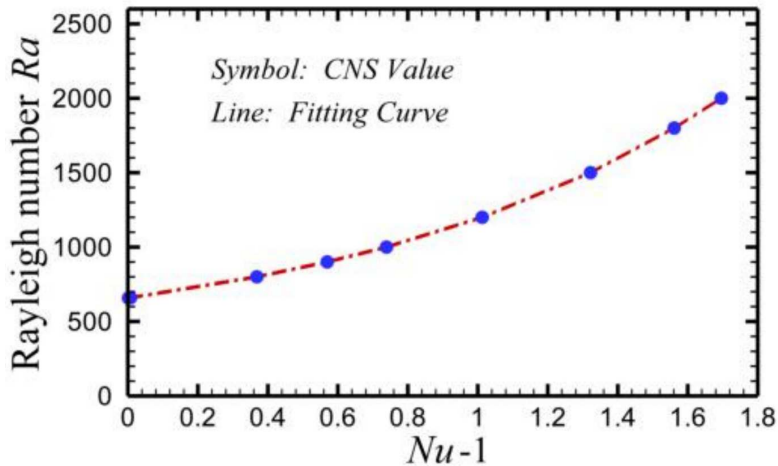}
        \end{tabular}
    \caption{{\bf Verification of CNS algorithm}. The Nusselt number $Nu$ is calculated by CNS (in symbol) with the double Fourier expansion modes $M = N = 31$ at different Rayleigh numbers above the critical value $Ra_c = 27\pi^4/4$. From the curve fitting (line) the estimated critical Rayleigh number agrees well with the theoretical value.} \label{VerifCNS}
    \end{center}
\end{figure}

To verify the correctness of our CNS algorithm, we first of all calculated the Nusselt number for several Rayleigh numbers a little larger than $Ra_c$ with $M=N=31$.  The results can be fit as
\begin{eqnarray}
Ra &=& 17.934 (Nu-1)^4 +52.599(Nu-1)^3\nonumber\\
&+& 131.01(Nu-1)^2 +330.66(Nu-1) \nonumber\\
&+& 657.46,
\end{eqnarray}
as shown in Fig.~\ref{VerifCNS}.  As $Nu\to 1$, the critical Rayleigh number can be estimated as $Ra_c \approx 657.46$, which agrees very well with the theoretical value $Ra_c = 657.5$.

\begin{figure}[t]
    \begin{center}
        \begin{tabular}{cc}
            \includegraphics[width=0.8\textwidth]{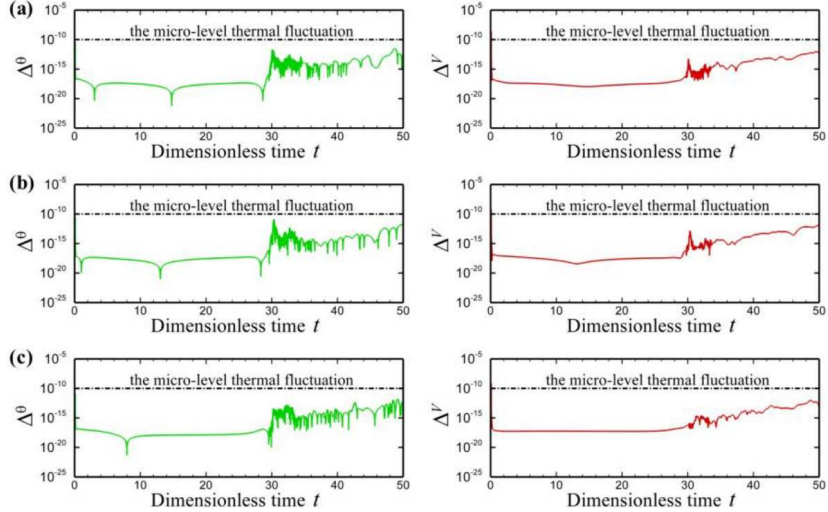}
        \end{tabular}
    \caption{{\bf Reliability check of the CNS results}. The results are for the case A  (of the initial thermal fluctuation) with the Rayleigh number $Ra = 10^{7}$ and the double Fourier expansion modes $M = N = 127$ at three probe points: ({\bf a}) ($3L/4, H/2$), ({\bf b}) ($3L/4, 2H/5$) and ({\bf c}) ($3L/4, H/10$).  The curves denote the dimensionless deviations of $\Delta^\theta_{10} = |\theta_{P=12}-\theta_{P=10}|/\theta_{RMS}$ ({\bf left}) and $\Delta^V_{10} = |V_{P=12}-V_{P=10}|/\sqrt{E_{RMS}}$ ({\bf right}), which are much less than the micro-level thermal fluctuation. Here $P$ is the order of the Taylor series in time; $\theta_{RMS}$ is the spatial root mean square of $\theta$ (the temperature departure from a linear variation background); $E_{RMS}$ is the spatial root mean square of the kinetic energy $(u^2+w^2)/2$, respectively.} \label{CNS}
    \end{center}
\end{figure}

Furthermore, to check the reliability of the CNS, the results from different orders of the Taylor series in time, e.g. $P=10$ and $P=12$, are compared at three probe points ($3L/4, H/10$), ($3L/4, 2H/5$) and ($3L/4, H/2$). Considering the butterfly-effect of the nonlinear dynamic system, the reliable results for the temperature $\theta$ (departure from a linear variation background) filed and the velocity field require that the deviations using the {\em same} initial condition must be much less than their respective spatial root mean square, i.e. $\theta_{RMS}(t)$ and $\sqrt{E_{RMS}(t)}$. As shown in Fig.~\ref{CNS}, at all probe points the nondimensionalized deviations are 10 orders of magnitude less than unity, while results from DNS are too largely deviated (15 orders of magnitude larger) to work adequately. Therefore, the CNS results obtained from the 10th-order Taylor series is reliable in the time interval $t\in[0,50]$.

\begin{figure}[t]
    \begin{center}
        \begin{tabular}{cc}
            \includegraphics[width=0.7\textwidth]{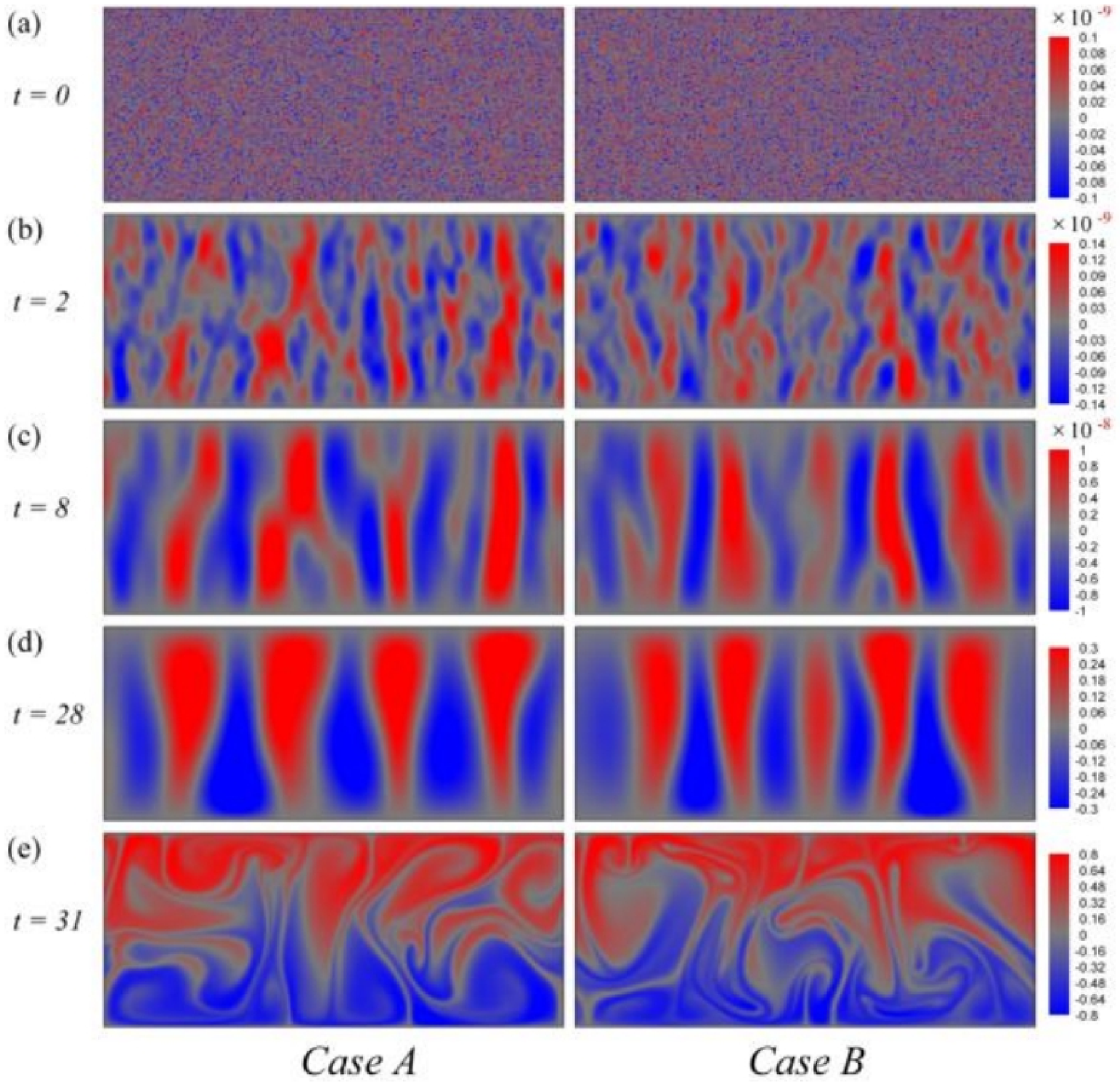}
        \end{tabular}
    \caption{{\bf Evolution of the $\theta$ (temperature departure from a linear variation background) field}. The results are for ({\bf a}) $t=0$; ({\bf b}) $t=2$; ({\bf c}) $t=8$; ({\bf d}) $t=28$ and ({\bf e}) $t=31$, with the Rayleigh number $Ra = 10^{7}$ and the double Fourier expansion modes $M = N = 127$. Case A and case B have different initial micro-level randomness, generated by the same variance of temperature $\sigma_T=10^{-10}$ and velocity $\sigma_u=10^{-9}$.} \label{structure}
    \end{center}
\end{figure}

\subsection{Evolution of the flow structure}

Although the RB convection is modeled theoretically as external disturbance isolated, randomness at the microscopic level still exists because of the molecular thermal fluctuation. In different CNS cases the initial temperature and velocity fields are randomly generated as Gaussian white noise, with the same temperature variance $\sigma_T=10^{-10}$ and velocity variance $\sigma_u=10^{-9}$, respectively. As shown in Fig.~\ref{structure} (a), such tiny difference of the initial condition is negligibly small with respect to the background fields at the macroscopic level, and thus the initial status can be regarded as the {\em same} from the physical viewpoint. With time increases, the filed structures and scales evolve rapidly. The clear large-scale patterns appear even at the very early stage, as shown in Fig.~\ref{structure} (b) for $t=2$, although the magnitudes are still insensibly small. In the following stage e.g. at $t=8$ as in Fig.~\ref{structure} (c) the large-scale structures become more and more distinct. Interestingly, these intermediate structures remain stable in a long interval up to $t=28$, as shown in Fig.~\ref{structure} (d), while the field energy increases continuously. At a critical point once the field is too energetic to be stable, these large-scale structures disintegrate abruptly, leading to the turbulent status as shown in Fig.~\ref{structure} (e) at $t=31$. Note that the two flow structures in Fig.~\ref{structure} (e) are sharply different, which must originate from the different initial microscopic randomness due to thermal fluctuation.

We emphasize that CNS can achieve the reliable results in a prescribed time interval with the numerical inaccuracy much less than the physical uncertainty, while DNS fails because of the butterfly effect. Therefore the evidence provided by CNS indicates that turbulence in the Reyleigh-B\'{e}nard convection problem can be self-excited or `out of nothing'~\cite{Tsinober2009}, i.e. the origin of randomness in fluid turbulence is intrinsic.

\begin{figure}[t]
    \begin{center}
        \begin{tabular}{cc}
            \includegraphics[width=0.6\textwidth]{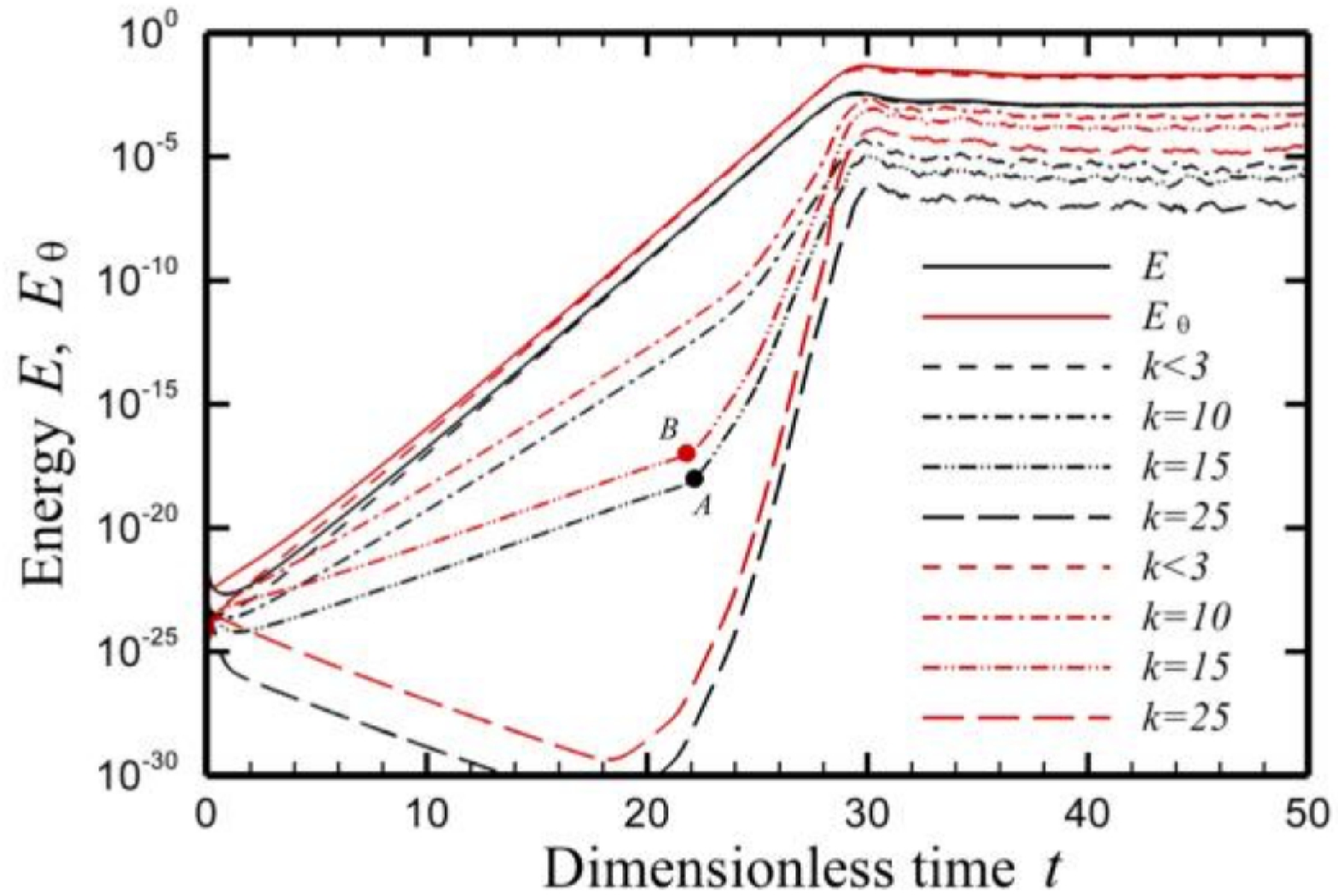}
        \end{tabular}
    \caption{{\bf Evolution of the kinetic and thermal energy at different scales}. The results are for case A with Rayleigh number $Ra = 10^{7}$ and the double Fourier expansion modes $M = N = 127$, where $E$ and $E_\theta$ denote the total kinetic and thermal energy, $k$ is the wave number, respectively. Lines in black: kinetic energy; Lines in red: thermal energy. The points $A$ (black dot) and $B$ (red dot) represent transition when the nonlinear interaction (with the large scale components) is strong enough to dominate to evolution process.} \label{energy}
    \end{center}
\end{figure}

\begin{figure}[t]
    \begin{center}
        \begin{tabular}{cc}
            \includegraphics[width=0.5\textwidth]{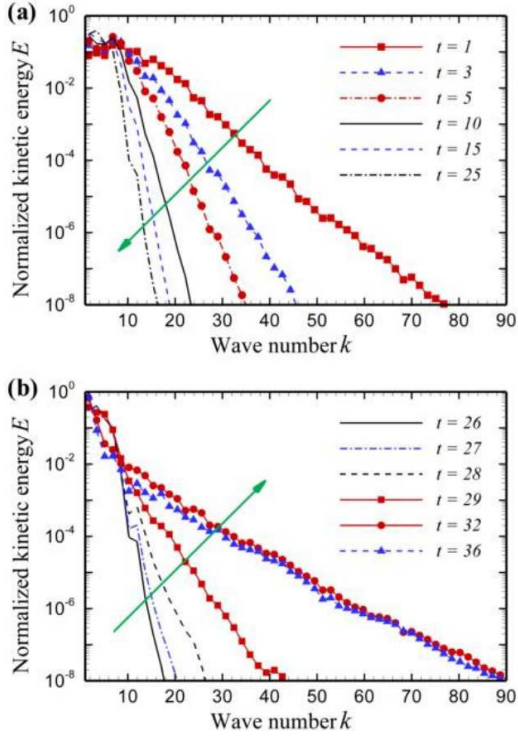}
        \end{tabular}
    \caption{{\bf Evolution of the normalized kinetic energy spectrum}. The results are for case A with Rayleigh number $Ra = 10^{7}$ and the double Fourier expansion modes $M = N = 127$. ({\bf a}) Initially energy shifts from small scales to larger scales, at which the spatial structure remains stable in most of the evolution process. ({\bf b}) when turbulence transition occurs the large scale disintegrates and energy shifts inversely from large scales to small ones.} \label{spectrum}
    \end{center}
\end{figure}

\begin{figure}
    \begin{center}
        \begin{tabular}{cc}
            \includegraphics[width=0.7\textwidth]{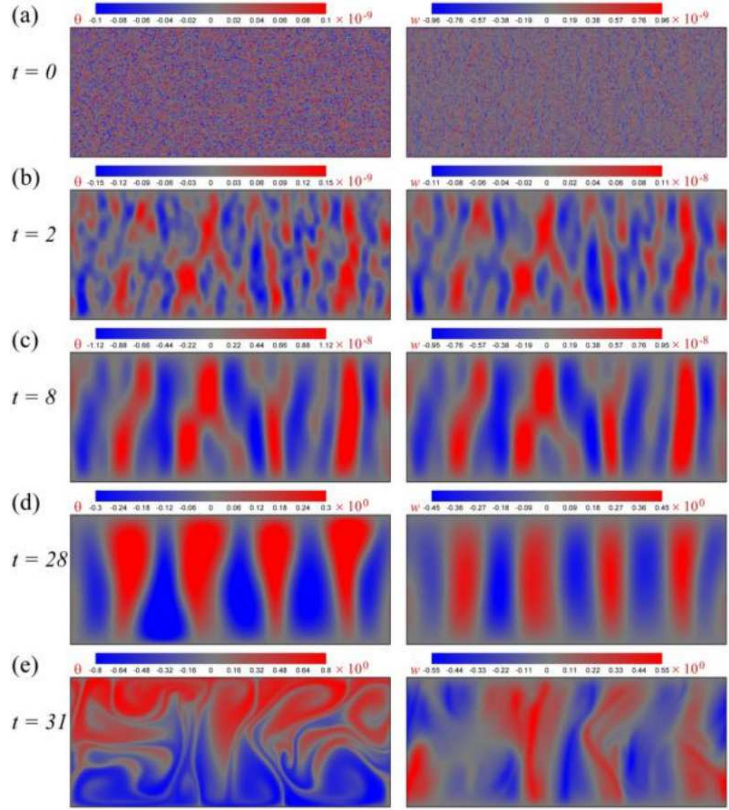}
        \end{tabular}
    \caption{{\bf Correlation between the $\theta$ field and $w$ field}. Here $\theta$ denotes the temperature departure from a linear variation background and $w$ is the velocity component along the opposite gravity direction, respectively. The results are for ({\bf a}) $t=0$; ({\bf b}) $t=2$; ({\bf c}) $t=8$; ({\bf d}) $t=28$ and ({\bf e}) $t=31$, with the Rayleigh number $Ra = 10^{7}$ and the double Fourier expansion modes $M = N = 127$.} \label{Correlation}
    \end{center}
\end{figure}

\begin{figure}
    \begin{center}
        \begin{tabular}{cc}
            \includegraphics[width=0.6\textwidth]{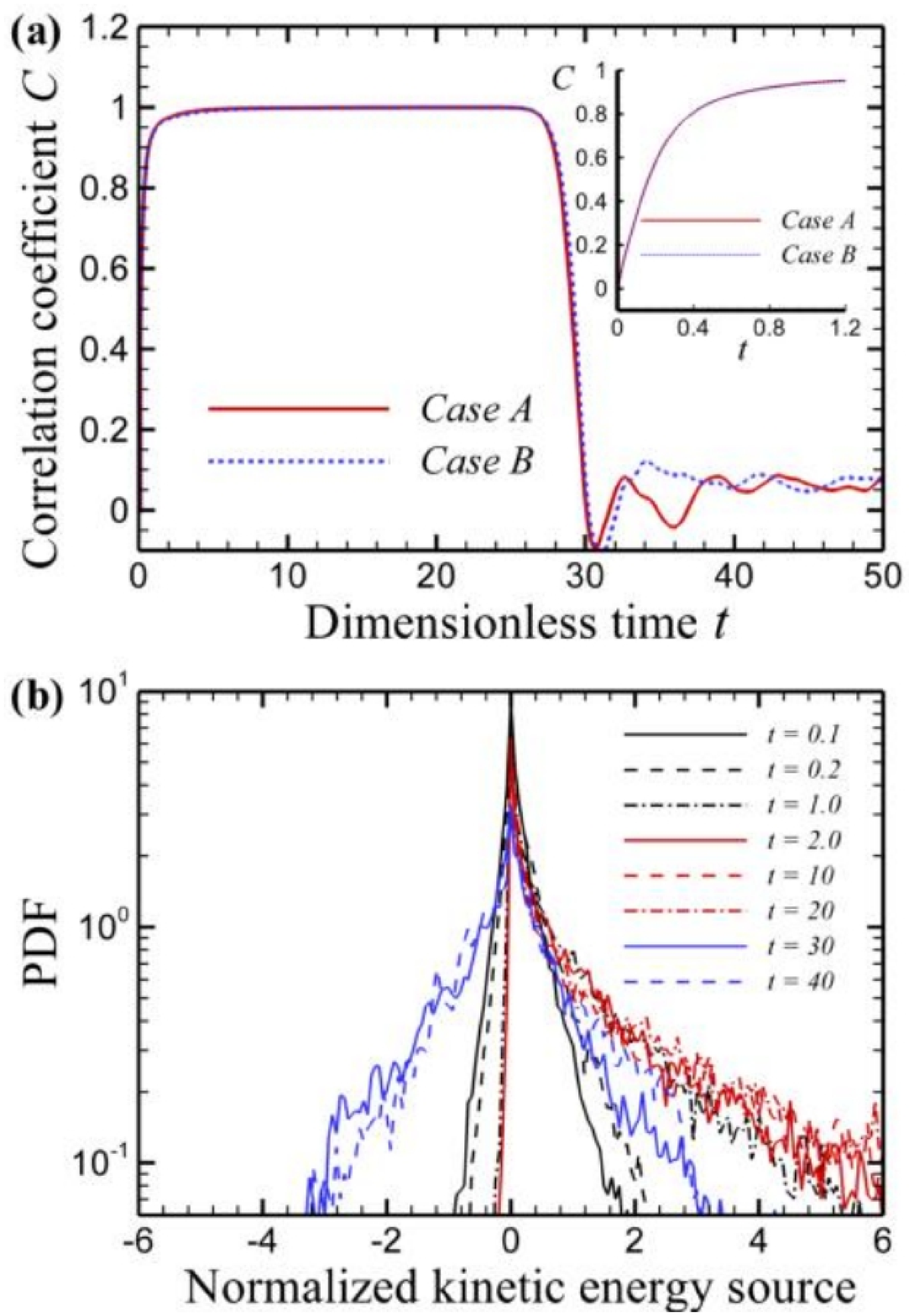}
        \end{tabular}
    \caption{{\bf Evolution of the correlation coefficient and its PDF}. ({\bf a}) Evolution of the correlation coefficient $C(t)$ between $w$  and $\theta$  for the case A with $Ra = 10^{7}$ and the double Fourier expansion modes  $M = N = 127$. From the initial state $C(t)$ increases rapidly (for $t<1$) from $C\sim 0$ because of the initial random independence to $C\sim 1$, corresponding to a strong correlation, which then remains till the transition to turbulence. ({\bf b}) Change of the PDF of the normalized kinetic energy source (for the case A with  $Ra = 10^{7}$). The PDF is initially close to be symmetric and evolve rapidly to be positively skewed. At the turbulence state, the PDF returns to be symmetric, but broadens largely.} \label{Evolution}
    \end{center}
\end{figure}

\subsection{Evolution of energy}

Considering the energy evolution, as shown in Fig.~\ref{energy}, both the total kinetic energy ($E$) and thermal energy ($E_\theta$) increase exponentially with identical slopes from the very beginning till the onset of turbulence at about $t = 31$, where a balance between energy absorption and dissipation is reached. For the individual components, the energy evolution is strongly dependent upon the wave number $k$. Most of the energy is contained in the large scale modes ($k<3$). As the wave number enlarges, energy increases exponentially first with a smaller growth rate at the beginning, but then increases superexponentially after a critical point, e.g. $A$ (or $B$) at about $t=22$ for $k=15$, till the onset of turbulence. For the mode with even larger wave number such as $k=25$, energy decays initially; and then increases after its critical point.

The energy change process can also be studied from the normalized kinetic energy spectrum. As shown in Fig.~\ref{spectrum}~(a), initially because of the strong influence from thermal fluctuation, the kinetic energy at higher wave number components decays rapidly so that spectrum recesses toward the large-scale side. Such large-scale dominant status remains till about $t=26$, when the spectrum begins to expand because more small scale modes are excited, as shown in Fig.~\ref{spectrum}~(b). During the unstable evolution process, the system gains energy from the background potential and restores most of the energy at the small wave number end.

Unstable evolution processes, including the RB convection, involve the following two interactions:
\begin{itemize}
\item Interaction between different scales (modes) due to nonlinearity.
\item Interaction between individual scales with the potential background, e.g. the mean temperature gradient in the RB convection.
\end{itemize}

\subsection{Channel of energy and randomness information transport}

Generally
Let $\dot{E}_{bg}(k)$,  $\dot{E}_{nl}(k)$ and $\dot{E}_{dissip}(k)$ denote for the wave number $k$ mode the growth rate of energy absorbed from the potential background, the energy growth rate due to the nonlinear interaction, and the energy dissipation rate, respectively. As commented by Landau~\cite{Landau}, the essence of transition to turbulence is an increase of the number of the excited modes (degrees of freedom), i.e.
\begin{equation}
\dot{E}_{bg}(k)+\dot{E}_{nl}(k) > \dot{E}_{dissip}(k)  \label{criterion}
\end{equation}
holds for large enough wave number $k$.

According to the instability theory \cite{Lin, Orszag1980}, {\em all} modes absorb energy exponentially from the potential background. Starting from the initial microscopic thermal randomness, scales are separated because of the system nonlinearity dispersion to generate smaller and larger scale components. At this stage, $\dot{E}_{nl}(k)$ is insignificant for all scales because of the tiny thermal fluctuation, i.e. $\dot{E}_{nl}(k) \approx 0$. For the large-scale modes ($k<3$), because energy dissipation is negligible, i.e. $\dot{E}_{dissip}(k) \approx 0$, Eq.~\eqref{criterion} always holds, which explains the exponential increase behavior in Fig.~\ref{energy}. As $k$ becomes larger, the energy dissipation is stronger to decrease the energy growth rate. If $k$ is even larger (such as $k=25$), too strong energy dissipation then leads to $\dot{E}_{bg}(k)+\dot{E}_{nl}(k)< \dot{E}_{dissip}(k)$, which explains the initial decay of the component energy, as shown in Fig.~\ref{energy} and Fig.~\ref{spectrum} (a).

As the system evolves to be more energetic, the nonlinear interaction part $\dot{E}_{nl}(k)$ becomes more important to transport energy from larger scales to smaller ones. The turning point $A$ (or $B$) in Fig.~\ref{energy} for $k=15$ indicates the dominance of $\dot{E}_{nl}(k)$. When the strong nonlinear interaction propagates towards the larger wave number components, more small scale modes are then excited, as shown in Fig.~\ref{spectrum} (b), which justifies the Landau's picture~\cite{Landau}.

Such background interaction $\mapsto$ nonlinear scale interaction $\mapsto$ scale dispersion scenario explains the structure and randomness evolution as well. As shown in Fig.~\ref{structure}, at the early stage the field changes rapidly to form a large-scale skeletal structure, which remains geometrically stable with continuous growth of the total kinetic and thermal energy till the transition to turbulence. Although initially the nonlinear part is negligibly small, it is still vital in information exchange in the following sense. Initial randomness at the microscopic level is transited to the large scale modes via the nonlinear interaction. Consequently such randomness information is inherited by the large scale modes, and survives with the evolution of these energy containing modes. When the nonlinear interaction is strong enough the structure information of the large scale modes can be transited back to small scale modes. This randomness transition mechanism may be important to understand ergodicity in turbulence at different scales. In summary the system nonlinearity in the unstable evolution process functions not only as instability excitation, but more as a channel to transport randomness information, and energy as well, from  microscopic to macroscopic level.

\begin{figure}
    \begin{center}
        \begin{tabular}{cc}
            \includegraphics[width=0.7\textwidth]{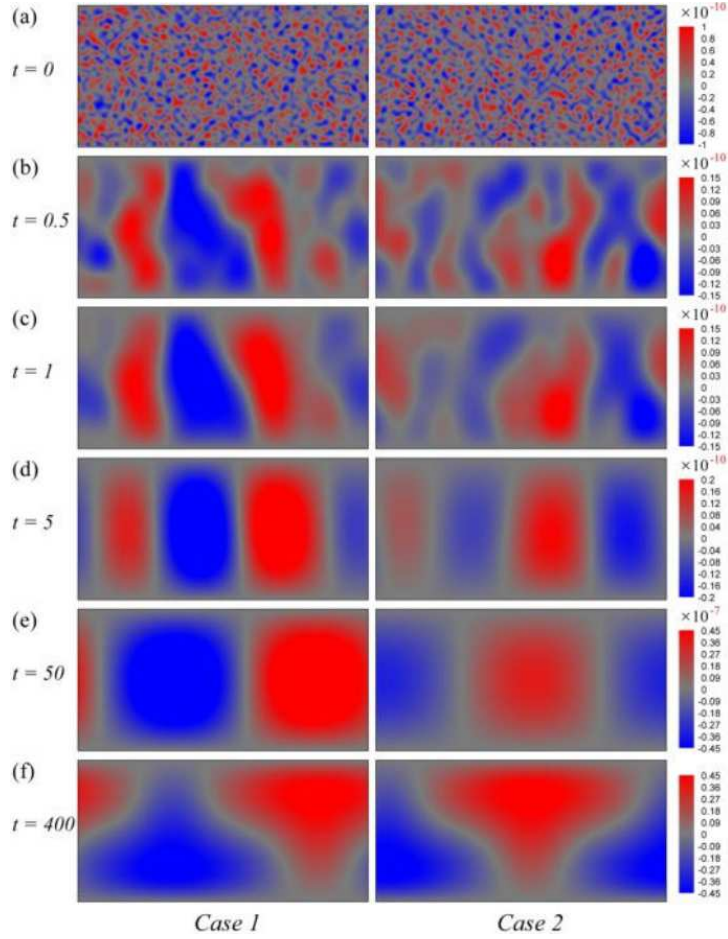}
        \end{tabular}
    \caption{{\bf Evolution of the $\theta$ (temperature departure from a linear variation background) field}. Here the Rayleigh number is $Ra = 2000$ and the double Fourier expansion modes $M = N = 31$. The results are for ({\bf a}) $t=0$; ({\bf b}) $t=0.5$; ({\bf c}) $t=1$; ({\bf d}) $t=5$; ({\bf e}) $t=50$; ({\bf f}) $t=400$. Case~1 and case~2 have different initial micro-level randomness due to thermal fluctuation, generated by the same variance of temperature $\sigma_T=10^{-10}$ and velocity  $\sigma_u=10^{-9}$.} \label{RaEq2000}
    \end{center}
\end{figure}

\begin{figure}
    \begin{center}
        \begin{tabular}{cc}
            \includegraphics[width=0.7\textwidth]{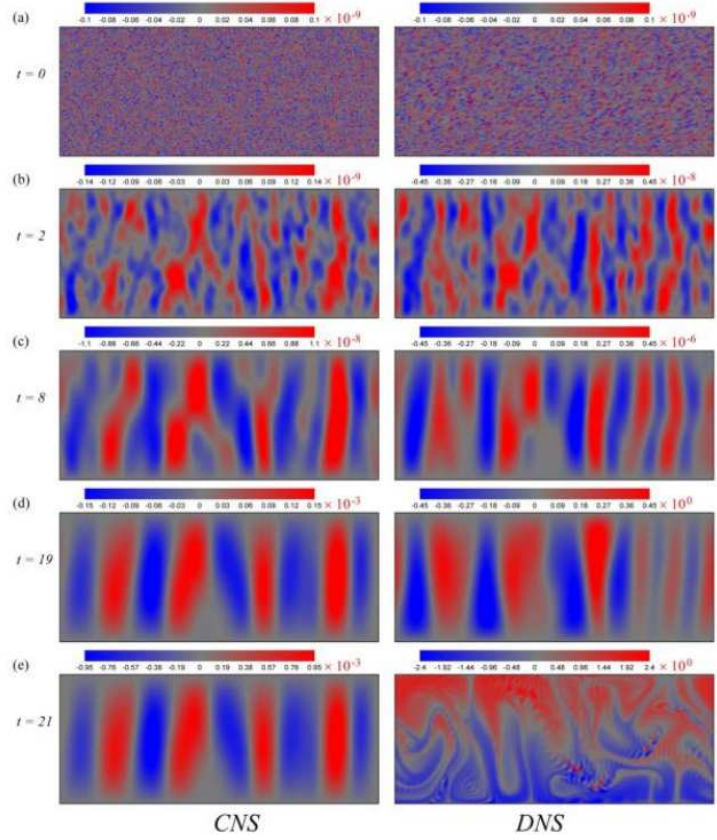}
        \end{tabular}
    \caption{{\bf Comparison of evolution of the $\theta$ (temperature departure from a linear variation background)  given by the CNS and DNS}.   Here, Rayleigh number is $Ra = 10^{7}$.   Left: CNS results, obtained using the same parameters in Figure 4 (Case A); Right: DNS results, obtained by means of the code DEDALUS using the resolution grid $M=N=127$,  the initial time step $dt = 0.005$, $cfl = 0.2$  and the same  initial guess as that of the CNS (Case A).} \label{comparison-CNS-DNS}
    \end{center}
\end{figure}

\begin{figure}[t]
    \begin{center}
        \begin{tabular}{cc}
            \includegraphics[width=0.7\textwidth]{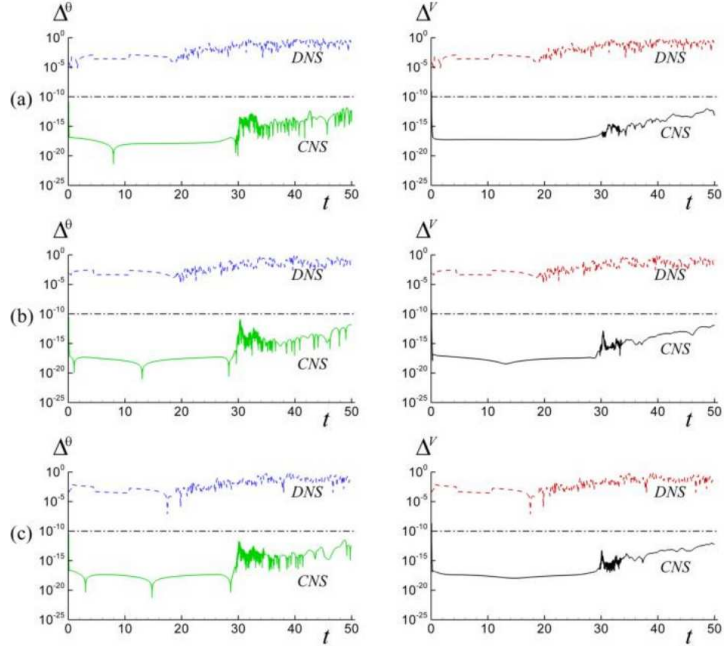}
        \end{tabular}
    \caption{{\bf Reliability check of the numerical results given by the CNS and DNS}.  1) for DNS (dash line)
$\Delta^\theta = |\theta_{cfl=0.2}-\theta_{cfl=0.1}|/\theta_{RMS}$
 and $\Delta^V = |V_{cfl=0.2}-V_{cfl=0.1}|/\sqrt{E_{RMS}}$; 2) for CNS (solid line) $\Delta^\theta = |\theta_{P=12}-\theta_{P=10}|/\theta_{RMS}$
 and $\Delta^V = |V_{P=12}-V_{P=10}|/\sqrt{E_{RMS}}$.  $\theta_{RMS}$ and  $E_{RMS}$ are the root mean squares of the temperature $\theta$ and  the kinetic energy $E=(u^2+w^2)/2$ in the domain $x\in[0,L], z\in[0,H]$. (a) At probe point ($3L/4, H/10$);
(b) At probe point ($3L/4, 2H/5$); (c) At probe point ($3L/4, H/2$).} \label{error-DNS}
    \end{center}
\end{figure}

\subsection{Some additional results}

Physically the correlation between the velocity and temperature field accounts for the source of the unstable evolution. Considering the kinetic energy budget, the source term is proportional to $w\theta$, where $w$ is the velocity component opposite to the gravity direction.  Fig.\ref{Correlation}  shows the individual field of $w$ and $\theta$  at different moments. Starting from the initial random states, these two fields evolve rapidly to be highly similar till the occurrence of turbulence.

The relation between these two field structures can be further quantified by the correlation coefficient $C(t)$ in the entire domain, as shown in Fig.~\ref{Evolution} ({\bf a}). Initially $C\sim 0$ because of the random independence. Very rapidly $C(t)$ increases almost to $1$ and remains invariant till turbulence occurs. The large value of $C(t)\sim1$ indicates that the $\theta$ field is almost perfectly correlated with the $w$ field. After transition to turbulence, $C(t)$ plunges and then fluctuates, but on average still remains above zero to balance the kinetic energy dissipation. Interestingly, the two $C(t)$ curves for different initial settings almost collapse till the onset of turbulence, as the  inherent micro-level uncertainty evolutes into the macroscopic randomness.

More details can be viewed from the probability density function (PDF) of $w\theta$, which is shown in Fig.\ref{Evolution} ({\bf b}), normalized by the overall mean of the instantaneous kinetic energy, i.e. $E_{RMS}(t)$. Initially, the part with negative $w\theta$ and the part with positive $w\theta$ are almost equal sized, which indicates the net contribution is close to be zero. However, in the evolution process the negative part shrinks rapidly and the PDF becomes strongly skewed toward the positive side, corresponding a significant net contribution from the source term. Once the flow transits to turbulent, the PDF approaches to be symmetrical again, but broadens largely, which corresponds to the strong macroscopic fluctuation inside the flow.

Moreover, it is also found from the present CNS results that the initial micro-level randomness can not be amplified when the Rayleigh number $Ra$ is under a critical value $Ra_c = 27\pi^4/4$. 
Even at $Ra = 2000 > Ra_c$, instability triggers the transition to a steady large-scale laminar flow {\em without} macroscopic randomness, as shown in Fig.~\ref{RaEq2000}. Therefore instability because of the system nonlinearity seems to be a necessary, but not sufficient, condition for the intrinsic macroscopic randomness evolution.

\section{Conclusion and discussions}

Following Lorenz~\cite{Lorenz1963}, we propose here a so-called `thermal fluctuation effect' to summarize the origin of intrinsic randomness in the Rayleigh-B{\'e}rnard convection system: a tornado can be created and its path can be ultimately altered due to intrinsic thermal fluctuation, {\em without} any external disturbances even from the wing flap of a butterfly. In methodology we also need to address the reliability of CNS, because the numerical noise can be well controlled even much lower than the microscopic thermal fluctuation. Although more expensive than DNS, CNS may open a new direction to understand the behaviors of turbulence.

Wolfram  \cite{Wolfram2002}  mentioned  that  the Lorenz equations with the famous butterfly-effect are highly simplified and thus do not contain terms that represent viscous effects.  So,  he believes that these terms would tend to damp out small perturbations.  However,  our CNS results  indicate that  the viscous effects of the NS equations can not  remove  its  sensitive dependance on the initial conditions (SDIC).   Thus, not only the Lorenz equation but also the full NS equations possess the property of the sensitivity dependance on initial conditions, i.e. the butterfly-effect,  which  implies  that the numerical noises should have a significant influence on numerical simulations of turbulence.

The open DNS code DEDALUS \cite{lecoanet2016}, which is available via http://dedalus-project.org/, is used to solve the same case (with Rayleigh number $Ra = 10^{7}$).      The  DNS solution  shows  very differently:  the transition to turbulence begins at about $t=19$, which is much earlier than the onset of turbulence given by the CNS, i.e.  $t  \approx 28$, as shown in Fig.~\ref{comparison-CNS-DNS}.   Besides, the DNS is strongly dependent upon control parameters (e.g. the time step).  It is easy to understand this numerical phenomenon,  since the numerical noises of the DNS are much larger than those of the CNS,  which (the numerical noises of the DNS) quickly transfer into the same order of magnitude as the background temperature field, as shown in Fig.~\ref{error-DNS}.  In other words,  due to the butterfly effect,  the numerical noises of the DNS themself become a large source of uncertainty.   In this sense the numerical uncertainty from DNS might be large enough to overwhelm the physical fidelity.    Strictly speaking,  the feasibility  of  DNS  on the non-equilibrium turbulence evolution still remains as an open question: current CNS results suggest that the numerical noises might  have a significant influence even on statistics of chaotic dynamic systems in non-equilibrium \cite{Li2016-B}.

In addition, we also use the CNS to solve the  Landau-Lifshitz Navier-Stokes (LLNS) equations \cite{Landau1959, Graham1974, Swift1977, Bell2010} of the same case (with Rayleigh number $Ra = 10^{7}$), where additional white noise fluxes are integrated into the N-S equations.  For the sake of brevity, mathematical details are omitted here.  The numerical results are qualitatively the same as those based on the NS equations.   Even quantitatively, they are close as well:  the onset of turbulence of the LLNS equation occurs at about $t=27$,  a  little  earlier  than  $t=28$  for  the  NS equation.  Physically, this is reasonable, since the thermal fluctuation always exists for the LLNS equations, but only exists at the beginning for the NS solution.  This suggests that the NS equations plus random initial condition due to thermal fluctuation could be a good approximation of the LLNS equations for the RB convection under consideration.
Note also that the thermal fluctuation propagates much less slowly  than the numerical noises.  So, it should be rather difficult to accurately simulate evolution of thermal fluctuation by means of the DNS.

All of these suggest that the CNS could provide us a new, more precise tool to investigate complicated nonlinear dynamic systems with sensitivity dependance on initial  conditions and numerical noises/algorithms,  for which significant interactions occur at different scales ranging from microscopic to macroscopic, although its wide applications might need a new generation computer in future.

\section*{Acknowledgment}
This work is partly supported by the National Natural Science of China (approval numbers 11272209 and 11432009). The parallel algorithms were performed on TH-1A at National Supercomputer Centre in Tianjin, China.  Thanks to Jing Li for using the open resource DEDALUS to gain the DNS results.

\end{document}